# ASTROPHYSICAL JETS AND OUTFLOWS

**Elisabete M. de Gouveia Dal Pino**
*Instituto Astronômico e Geofísico, Universidadede São Paulo (IAG-USP)*
*dalpino@astro.iag.usp.br*

## ABSTRACT

Highly collimated supersonic jets and less collimated outflows are observed to emerge from a wide variety of astrophysical objects. They are seen in young stellar objects (YSOs), proto-planetary nebulae, compact objects (like galactic black holes or microquasars, and X-ray binary stars), and in the nuclei of active galaxies (AGNs). Despite their different physical scales (in size, velocity, and amount of energy transported), they have strong morphological similarities. What physics do they share? These systems are either hydrodynamic or magnetohydrodynamic (MHD) in nature and are, as such, governed by non-linear equations. While theoretical models helped us to understand the basic physics of these objects, numerical simulations have been allowing us to go beyond the one-dimensional, steady-state approach extracting vital information. In this lecture, the formation, structure, and evolution of the jets are reviewed with the help of observational information, MHD and purely hydrodynamical modeling, and numerical simulations. Possible applications of the models particularly to YSOs and AGN jets are addressed.

## 1. INTRODUCTION

Astrophysical jets are physical conduits along which mass, momentum, energy and magnetic flux are channeled from stellar, galactic and extragalactic objects to the outer medium. Geometrically, these jets are narrow (small opening angle) conical or cylindrical/semi-cylindrical protrusions (e.g., Das 1999). Jets are an ubiquitous phenomenon in the universe. They span a large range of luminosity and degree of collimation, from the most powerful examples observed to emerge from the nuclei of active galaxies (or AGNs) to the jets associated to low-mass young stellar objects (YSOs) within our own Galaxy. In the intermediate scales between these two extremes one finds evidences of outflows associated to neutron stars, massive X-ray binary systems (with SS433 being the best example of this class), symbiotic stars, and galactic stellar mass black holes (or microquasars).

Less collimated supersonic outflows are often observed to emerge from massive hot stars in their late stages of stellar evolution, like the LBV (luminous-blue-variable) stars (with eta-carinae being the most powerful object of this class) and also from low mass stars in the late stages, like the proto-planetary nebulae (see, e.g., Gonzalez et al. 2004, Soker 2004, and references therein).

Also apparently associated to jet phenomena are the gamma ray bursts (GRBs). Discovered nearly 30 years ago, at cosmological distances they seem to be the most powerful sources in the Universe (with luminosities $10^{51-53}$ erg/s). There is now some evidence that they are emitted from relativistic jets via synchrotron emission. Some GRBs (long bursts) seem to be associated with supernovae (Mészáros 2002).

Although out of the scope of the present lecture, it is also worth noting that recent space observations of our Sun, have revealed that the solar corona is full of jets and flares. Though the total energy of the solar jets and flares is much smaller than those in comic jets, the spectrum and time variability of electromagnetic waves emitted from solar flares are quite similar to those of cosmic flares (e.g., Shibata 2003).

As largely stressed in the literature, most of these outflows, despite their different physical scales and power, are morphologically very similar, suggesting a common physical origin (see below). For example, in one extreme, AGN jets have typical sizes $\geq 10^6$ pc [1], nuclear velocities $\sim c$ (where c is the light speed), and parent sources (which are massive black holes) with masses $10^{6-9}$ $M_\odot$ [2] and luminosities $\sim 10^{43-48}$ $L_\odot$ [3]; while in the other extreme, YSO jets have typical sizes $\leq 1$ pc, nuclear velocities $\leq 10^{-3}c$, and emerge from low mass protostars with masses $\sim 1$ $M_\odot$. and luminosities $(0.1 - 2 \times 10^4)$ $L_\odot$. Therefore, the jet phenomenon is seen on scales that cover more than seven orders of magnitude. Nonetheless, all the jet classes share common properties. In general, they: (i) are highly collimated and in most cases two-sided; (ii) originate in compact objects; (iii) show a chain of more or less regularly spaced emission knots which in some cases move at high speeds away from the central source; (iv) often terminate in emission lobes (with line emission in the case of the YSOs and synchrotron continuum emission in the case of the AGN jets), which are believed to be the "working surfaces" where the jets shock against the ambient medium (see below); (v) are associated with magnetic fields whose projected directions are inferred from polarization measurements; and (vi) show evidence of accretion of matter onto the central source via an accretion disk (e.g., Königl 1986, de Gouveia Dal Pino 1995). For comparison, Figs. 1 and 2 show some of the finest examples of YSO and AGN jets.

---

[1] 1 pc = 1 parsec = 3.086 $10^{18}$ cm.

[2] 1 $M_\odot$ = one solar mass = 1.99 $10^{33}$ g

[3] 1 $L_\odot$ = one solar luminosity unit = 3.826 $10^{33}$ erg/s

Recent reviews of the observational and structural properties of the YSO jets can be found in Reipurth and Bally (2001, see also de Gouveia Dal Pino 1995, Reipurth 1999, de Gouveia Dal Pino and Cerqueira 2002, Raga et al. 2003, de Gouveia Dal Pino et al. 2003, Cerqueira and de Gouveia Dal Pino 2003). Reviews of the properties of the AGN jets and the relativistic galactic jets from X-ray binaries and microquasars can be found in, e.g., Das (1999), Shibata (2003), Pringle (1993), Livio (1998, 2002), Mirabel and Rodriguez (1999, 2003), and Massaglia (2003).

In this work, instead of discussing all the current knowledge about the various classes of astrophysical jets (which would be impossible to cover just in few pages), I focus mainly on the YSO jets, which are the closest and for this reason, excellent laboratories for cosmic jet investigation, and on AGNs jets. Some specific issues about the theory of microquasar jets will be also addressed.

The review begins with a brief description of the observed characteristics of the YSO, AGN and microquasar jets (Section 2). A discussion of the basic structural components of the jet and the analytical and numerical models which explain these features is presented in Sections 3. All sort of jet propagation effects like momentum transfer to the environment, jet precession, and deceleration, are addressed in Session 4. The effects of magnetic fields on the jet evolution and structure are discussed in Session 5, and finally, in Session 6, the existing mechanisms for the origin and collimation of the jets from the central object and associated accretion disk is reviewed.

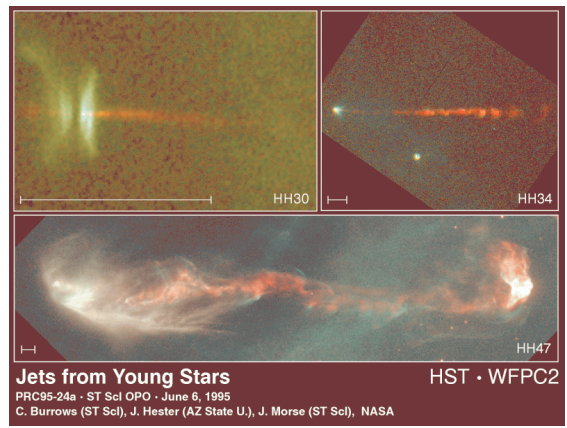

**Figure 1.** Hubble Space telescope images of three YSO (also called Herbig-Haro, or simply HH) jets. The HH 30 jet is observed to emerge from the embedded source surrounded by a disk of gas and dust. Terminal bright bow shocks are clearly seen on both sides of the HH46 jet, and a chain of emission knots in the HH34 jet. The white line in each frame is 1000 AU or about the scale of our solar system (extracted from the HST website).

## 2. OBSERVATIONAL PROPERTIES OF YSO AND AGN JETS

### 2.1. YSO Jets

Protostellar (YSO) jets are produced during the major accretion phase in the star-formation process, i.e., during the Class 0 and I phases of the life of a protostellar core/young star. This phase of the star formation process is thought to last about $10^5$ yr (Lada 1999).

The YSO jets have typical projected lengths between ~0.01 - few pcs. Most of them show a linear chain of bright, traveling knots, frequently identified as Herbig-Haro (HH) objects (e.g., HH34, and HH111 systems). They often terminate in a bow shock-like structure identified as the leading *working surface* where the jet impacts with a slower ambient gas (Fig. 1). Researchers at first considered the most luminous Herbig-Haro object in the flow to be the terminal bow shock, i.e., to mark the location where the proto-stellar flow impacts its cloud environment. However, these regions are typically displaced only about 0.1 pc from the star. With the recent availability of wide-field CCD arrays, it has become clear that some jets indeed extend to far greater distances showing two or more bow shock structures separated by a trail of diffuse gas for many jet radii. (Devine et al. 1997). Examples of these *giant HH flows* include HH111 with a total extent of ~ 7.7 pc, HH34 of ~ 3 pc, and HH355 with a total extent of ~ 1.55 pc.

In some cases (e.g., HH30 jet), there is no detected bow-shock like feature (Raga 1993). Other jets, instead of chains of aligned knots, show larger amplitude side-to-side "wiggles" (e.g., HH83 and HH110 jets; Reipurth 1989).

All the luminous structures produce emission line spectra in the optical and infrared bands mainly. Spectral lines give information on local temperature and density, on the bulk velocity of the jet emitting matter and on the presence of shocks along the jet. Prominent features include the hydrogen Balmer lines and transitions of neutral atoms ([OI], [CI], [NI]) and ions (CaII, [CaII], [FeII], [SII], [OIII]). While the terminal feature may be rich in high ionization and excitation lines, the inner jet generally shows a very low excitation spectrum with [SII] and [OII] to H$\alpha$ ratios substantially greater than unity. This implies that the temperatures of the YSO jets are not much larger than $\approx$ 1-2x$10^4$ K and the corresponding sound speeds are ~10 km/s. This yields typical Mach numbers (the ratio of the jet speed to the sound speed) for the emitting regions $M_j$=20-40 (Raga 1993).

From the Doppler shifts of the emission lines and proper motions it is possible to constrain the jet velocity. The knots move away from the sources at speeds ~100-500 km/s, have radii $R_j \approx 3 \times 10^{15}$ cm, and inter-knot separation $\Delta x \approx 10^{16}$ cm $\approx$ 3.3 Rj.

Typical electron densities (obtained from the assumption that the line emission is produced by shock-heated gas) range from $n_j \approx 10$ cm$^{-3}$ for the faintest objects to $> 10^5$ cm$^{-3}$ for the brightest (Reipurth and Bally 2001). An increasing angular resolution and sensitivity achievable in recent years with the large telescopes has allowed to investigate the velocity structure across the jet and revealed the jet characteristics close to the central source. In these studies, an average hydrogen jet number density $10^4$ cm$^{-3}$ has been obtained (Bacciotti and Eislöffel 1999).

The high proper motion of the heads of the jets indicate that their density is considerably higher than the density of the surrounding medium. Although uncertain, observations suggest a jet-to-environment density ratio $\eta = \rho_j/\rho_a \approx$ 1-20 (e.g., Morse et al. 1992, Reipurth and Bally 2001).

### 2.2. AGN jets

These are observed to emerge from the nuclei of active galaxies, like Seyfert galaxies, distant quasars, and radio galaxies, and may extend for distances up to few mega-parsecs into the intergalactic medium. Some of them are considered to be the largest single coherent structures found in the Universe.

None of the basic parameters, like the jet velocity, the Mach number, or the jet to the ambient density ratio can be directly constrained by observations in the case of AGN jets. Observers must therefore interpret their data relying upon statistical analyzes and at an assumed basic model.

In contrast to YSO jets, the main difficulty one faces when trying to comprehend the nature of jets from Active Galactic Nuclei (and also the jets from galactic black holes) is the absence of lines in the radiation spectrum of these objects. Their emission is typically continuous and non-thermal in a wide frequency range that goes from the radio to the X-ray bands (and is due to synchrotron and inverse Compton emission[4]). This is basically the reason why after almost five decades since the discovery of the radio emission from the jet in the galaxy Cygnus A (Jennison and Das Gupta 1953) observers and theorists are still debating about very basic questions such as the extragalactic jets composition, i.e., whether they are made of ordinary matter (electron-proton) or of electron-positron pairs (e.g., Massaglia 2003). On the other hand, the observation of synchrotron radiation is an evidence of the presence of both charged particles (possibly mostly electrons moving away from the source at relativistic velocities) and magnetic fields.

---

[4] Whenever a relativistic charged particle encounters a magnetic field line it spirals around it and emit polarized *synchrotron* radiation at meter and centimeter (radio) wavelengths. Since the electrons are much lighter and therefore, much faster moving than, e.g., the protons, they are responsible for most of the observed synchrotron radiation from the AGN jets. Inverse Compton is the radiation that is produced (generally at X-rays wavelengths) by photons from the environment which are scattered by relativistic electrons to higher frequencies. These mechanisms are both non-thermal in nature because there is no link between the emission they produce and the temperature of the radiating object. Their radiation spectrum is therefore, not described by a black-body curve.

At parsec scales, close to the source, several AGN jets exhibit a series of bright components (or knots) which travel from the core with apparent superluminal motion (i.e., a velocity apparently greater than the light speed). This apparent effect is interpreted as a consequence of relativistic motions in a jet propagating at a small angle to the line of sight with flow velocity as large as ~ 99% of the speed of light (and in some cases, even beyond). The same phenomenon is detected in galactic microquasars. These inferred jet velocities close to the speed of light suggest that jets are formed within a few gravitational radii of the event horizon of the black hole.

Based on their morphology at kilo-parsec scales, the extragalactic jets observed in radio galaxies have been historically classified into two categories (Fanaroff and Riley 1974): a first class of objects, preferentially found in rich clusters of galaxies and hosted by weak-lined galaxies, shows jet-dominated emission and two-sided jets and was named FR I (Fig. 2, top); a second one, found isolated or in poor groups and hosted by strong emission-line galaxies, presents lobe-dominated emission and one-sided jets and was called FR II (or "classical doubles") (Fig. 2, bottom).

Apparently, the morphology and dynamics of the jets at kilo-parsec scales are dominated by the interaction of the jet with the surrounding extragalactic medium that tends to decelerate the flows to non-relativistic velocities, although some jets (FRII) seem to remain relativistic even on these large scales (see Session 4).

A clear indication of supersonic speeds in these jets at kilo-parsec scales, is the presence of shocks that are resolved in the transverse direction, and the signature of shocks is given by the behavior of the polarization vector of the synchrotron radiation. Well-resolved jets are often highly linearly polarized and their magnetic fields are thus partially ordered. Detailed polarization maps obtained in the optical and radio bands of the jets of M 87 (the closest AGN) (Perlman et al. 1999), for example, show that the magnetic field is mainly parallel to the jet axis but becomes predominantly perpendicular in the regions of emission knots, indicating a shock compression of the field lines along the shock front. Polarization can be also detected in the prominent radio lobes and hot spots that form in some jets at the region where they impact with the intergalactic medium. Again, the field direction is oriented along the lobe's border, i.e., perpendicular to the jet axis indicating compression by shock.

Another parameter to constrain is the jet-to-ambient density ratio. For this parameter one has to rely upon analytical and numerical modeling and compare qualitatively the spatial density distribution, resulting from the calculations, with the observed brightness distributions. Simulations of supersonic, underdense jets are able to reproduce the overall picture of the observed FRII jets (e.g., Massaglia, Bodo, and Ferrari 1996). They are, therefore, *underdense* with respect to the ambient medium, with $\eta \sim 10^{-5}$ in the case of pair plasma (proton+electron) jets (Massaglia 2003).

We notice that the basic reason why shocks lead to non-thermal radiation in AGN jets and thermal emission in YSO jets is the 6 to 8 orders of magnitude ratio in the ambient density. YSO jets propagate in a high density ambient medium of molecular clouds and their shocks can, therefore, heat the high density matter that will in turn suffer efficient radiative losses (which are $\propto \rho^2$, where $\rho$ is the gas density) through line emission. On the other hand, extragalactic jets propagate in a tenuous intergalactic or intracluster medium and the shocks can be considered with good approximation *adiabatic* (see Session 3). They will have the main effect of accelerating the particles to relativistic velocities via a first-order-Fermi process[5] that, in turn, will yield synchrotron radiation in the shocked ambient magnetic field and the only possible signature that we observe of these shocks is the reorientation of the polarization vector of the radiation.

### 2.3. A note on Microquasar Jets

Microquasar or relativistic jets from stellar-mass black holes in binary stars emitting X-rays (also denominated BHXRTs), are scaled-down versions of AGN (or quasar) jets, typically extending for ~ 1pc and probably powered by spinning black holes with masses of up to few tens of the Sun. Despite the enormous difference in scale, both classes share a lot of similarities in their physical properties. They are both believed to be surrounded by an accretion disk (see Session 6) and since the characteristic times in the flow of matter onto the black hole are proportional to its mass, the accretion-ejection phenomena in microquasars is expected to last much shorter, about $10^{-7}$-$10^{-5}$ faster than analogous phenomena in quasars. For this reason they are easier to investigate - variations on scales of ten minutes of duration in microquasars, for example, would have been difficult to observe in quasars. For a review on their observational properties see Mirabel and Rodriguez (1999).

---

[5] First-order Fermi acceleration: when charged particles encounter a randomly moving region with high magnetic field the particles energy will increase due to head-on-collisions with the magnetic fluctuations. In the radio jets this process occurs right behind the shock fronts where the magnetic fields are amplified by compression. The electrons which are accelerated by this process emit synchrotron radiation.

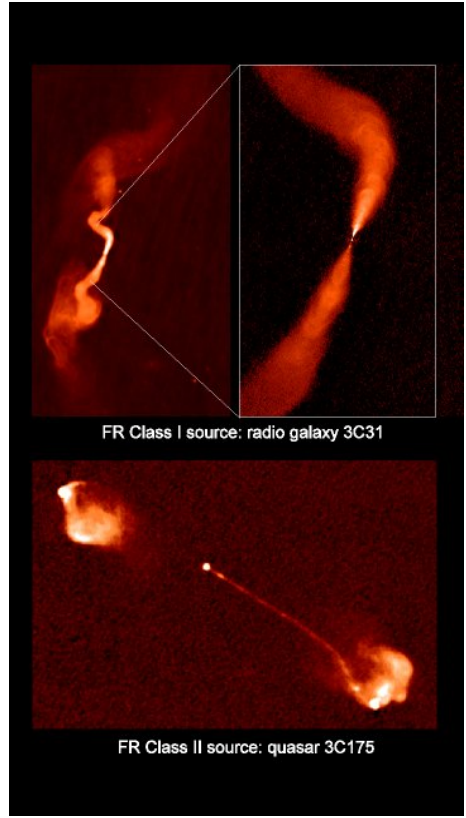

**Figure 2.** Images obtained with the Very Large Array (VLA) of jets in a FRI I source (top: 3C31 at the radio frequencies 1.4 GHz and 8.4 GHz), and a FR II source (bottom: 3C175 at 4.9 GHz); extracted from Bridle (1998).

## 3. BASIC JET STRUCTURE

### 3.1. Jet Head

A supersonic jet propagating into a stationary ambient gas will develop a *double* shock pattern at its head (also denominated *working surface*) due to impact with the ambient gas - while the impacted ambient material is accelerated by a forward bow shaped shock, the beam outflow is decelerated in a jet shock or Mach disk. The velocity of advance of the bow shock can be estimated by balancing the momentum flux of the jet material at the jet head with the momentum flux generated by the ambient medium at the bow shock:

$$[\rho_j (v_j - v_{bs})^2 + p_j] R_j^2 \approx [\rho_a v_{bs}^2 + p_a] R_h^2 \quad (1)$$

For highly supersonic flows, the thermal pressure of the jet ($p_j$) and the ambient medium ($p_a$) can be neglected and one obtains (e.g., de Gouveia and Benz 1993):

$$v_{bs} \approx v_j [1 + (\eta\alpha)^{-1/2}]^{-1} \quad (2)$$

where $\rho_j$ and $\rho_a$ are the jet and ambient densities, respectively, $v_j$ is the jet velocity, $v_{bs}$ is the velocity of advance of the bow shock, $\eta = n_j/n_a$ is the ratio of the jet number density to the ambient number density, and $\alpha = (R_j/R_h)^2$, where $R_j$ is the radius of the jet beam and $R_h$ is the radius at the jet head.

Eq. (2) shows that for a low-density jet ($\eta \ll 1$) which is generally believed to be the case for AGN jets, $v_{bs} \ll v_j$ and the jet material is constantly decelerated at the end of the jet (at the jet shock). The high-pressure, shocked gas then drives a flow backward around the jet forming a *cocoon* of shock-heated waste material (see Fig. 2).

A dense jet, like typical YSO jets, on the other hand, for which $v_{bs} \approx v_j$ ($\eta >> 1$), will simply plow through the ambient medium close to the jet velocity without accumulating much waste of gas in the cocoon. The ambient material that traverses the bow shock will form a *shroud* of ambient shock-heated gas surrounding the beam/cocoon structure. Fig. 3 illustrates the basic components of a supersonic jet.

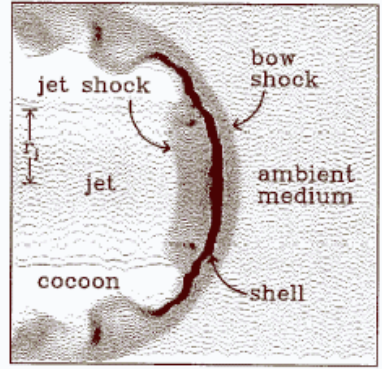

**Figure. 3.** The basic structural features of a jet: the beam, the working surface (with the bow shock and the jet shock), the cocoon, and the surrounding shroud (from Blondin, Konigl, and Fryxell 1989).

Supersonic flows occurring in AGNs were the focus of much theoretical work since their discovery. Extensive numerical simulations were performed assuming *adiabatic* jets in order to study their structure, stability properties, and propagation through the ambient medium (Norman 1990, Bridle 1998, Das 1999). As remarked before, AGN jets generally evolve dynamically like *adiabatic* flows.

In the case of the YSO jets, however, it has been known for quite a while that the dynamics of these jets is very much affected by *radiative cooling* due to recombination of the shock-heated gas (Blondin, Fryxell, and Konigl 1990, de Gouveia Dal Pino and Benz 1993, Stone and Norman 1993, Chernin et al. 1994). The radiative cooling time of the shock-heated material in a thermal flow is given by:

$$t_{cool} = (n_e + n_H) k T_s / [(\gamma - 1) n_e n_H \Lambda(T_s)] \qquad (3)$$

where $n_e n_H \Lambda(T_s)$ is the cooling rate in units of energy per unit volume and time, $n_e$ and $n_H$ are the postshock electronic and hydrogen number densities, respectively, $\gamma$ is the ratio of specific heats of the gas, k is the Boltzmann constant, and $T_s$ is the immediate postshock temperature $T_s = 2(\gamma-1) \mu v_s^2 / [(\gamma+1) 2 k]$, where $\mu$ is the average mass per particle. Taking shock velocities $v_s \sim 40$-$100$ km/s, inferred from the observed line-ratios of the emitting matter in the knots, and number densities $\approx 100$ cm$^{-3}$ one finds $t_{cool} \approx 100$ yrs, which is much smaller than the evolutionary time of these YSO jets, $t_{dyn} = l_j/v_j \sim 1000$-$10,000$ yrs (where $l_j$ is the jet extension). The assumption of an adiabatic gas is, therefore, inappropriate for YSO jets.

The analytic predictions of the equations above and the importance of the cooling on the YSO jet dynamics have been confirmed by multidimensional hydrodynamical simulations (see session below) (e.g., Blondin, Fryxell, and Konigl 1990, de Gouveia Dal Pino and Benz 1993, Stone and Norman 1993; see also Reipurth and Raga 1999; Cabrit, Raga, and Gueth 1997 for reviews). In a radiative cooling jet, a dense shell develops in the head due to the cooling of the shock-heated gas (see Fig. 8, top panel). The shell is responsible for most of the emission of the jet. In Figure 8, the shell has become convectively (Rayleigh-Taylor) unstable and fragmented into pieces that have spilled out to the cocoon forming, together with the shell, an elongated plug of cold gas in the head. The fragmented shell resembles the clumpy structure observed in many HH objects at the bow shock of YSO jets (e.g., HH1, HH2, HH19, HH12).

### 3.2. Why Numerical Simulations?

While theoretical models are useful to understand the basic physics underlying these objects, numerical simulations allow us to go beyond the one-dimensional, steady-state treatment and disentangle the jet structure and evolution. The conservation equations that can describe the dynamical evolution of a magnetized *non-relativistic* flow in the absence of dissipation effects are the ideal magnetohydrodynamic (MHD) equations:

$$\frac{d\rho}{dt} = -\rho \nabla \cdot v,$$

$$\frac{dv}{dt} = -\frac{\nabla p}{\rho} + \frac{1}{4\pi\rho}(\nabla \times B) \times B,$$

$$\frac{du}{dt} = -\frac{p}{\rho}(\nabla \cdot v) - \mathscr{L},$$

$$\frac{dB}{dt} = -B(\nabla \cdot v) + (B \cdot \nabla)v,$$
(4)

where the symbols have their usual meaning, i.e., $\rho$ is the density; $B$ is the magnetic field; $u$ is the specific internal energy, $\mathscr{L}$ is the radiative cooling rate, $p$ is the gas pressure and $v$ is the flow velocity. To close this system, an equation of state is required and in the case of an ideal gas:

$$p = (\gamma - 1)\rho u,$$
(5)

where $\gamma$ is the ratio of specific rates of the gas. In the set of equations (4) above, the first one gives the mass conservation, the second the momentum conservation, the third the energy conservation, and the last one is the magnetic induction equation in its ideal form, without the magnetic resistivity term. The numerical integration of these equations in time and space in one, two, or three dimensions taking appropriate initial conditions allows one to trace the historical evolution of a flow.

Models that consider the jet overall dynamics, depend on a minimum of three parameters: the jet velocity, the Mach number, and the jet-to-ambient density ratio. Another crucial parameter for jet modeling is the magnetic field intensity which under some circumstances can be neglected - in some cases, it is possible to assume that the bulk kinetic energy density in the jet dominates on the magnetic energy and the general behavior of the jet propagation can then be captured by a pure *hydrodynamic* model (see below).

### 3.3. The origin of the bright knots

Most of the available models to explain their formation rely on the belief that the internal knots in the jets represent shocks that are excited within the jet flow.

*3.3.1. YSO jets*

Among the proposed models, Kelvin-Helmholtz (K-H) shear instabilities at the boundary between the jet and the surrounding medium could excite a pattern of oblique internal shocks (e.g., Ferrari, Massaglia, and Trussoni 1982, Birkinshaw 1997). This mechanism produces shocks that travel at a velocity $v_p$ that can be substantially smaller than the fluid velocity $v_j$. Shocks driven by this instability are viable candidates for knot production in jets for which the adiabatic approach is good, i.e., AGN or galactic microquasar jets. For radiative cooling jets (YSO jets), however, the cooling reduces the pressure in the cocoon that surrounds the beam. As a result, the cocoon has less pressure to drive K-H instabilities and thus, reflect internal shocks in the beam. In numerical simulations of radiative jets propagating continually into homogeneous ambient medium, only very *heavy* (overdense) jets ($\eta \approx 10$) show the formation of internal shocks (e.g. de Gouveia Dal Pino and Benz 1993). However, an efficient formation of a "wiggling" chain of weak knots by K-H instabilities, primarily close to the jet head, is found in simulations of jets propagating into environments of increasing density (and pressure) (de Gouveia Dal Pino and Birkinshaw 1996, de Gouveia Dal Pino, Birkinshaw, and Benz 1996). In these cases the cocoon is compressed by the ambient ram pressure and the beam is highly collimated. The compressing medium then induces the development of the K-H instabilities. This structure remarkably resembles some observed YSO jets (e.g., HH 30 jet) that show a wiggling structure with knot formation closer to the jet head.

The aligned knots observed in typical YSO jets, which frequently exhibit a bow shock like morphology and high spatial velocity (Reipurth and Bally 2001) are now believed to be mainly formed by *time-variability in the flow ejection velocity*. Supersonic velocity variations in the underlying outflow (with periods close to the transverse dynamical timescale $\tau_{dy} = R_j/v_j$) quickly evolve to form a chain of regularly spaced radiative shocks along the jet with large proper motions and low intensity spectra in agreement with the observed properties of the knots. The knots widen and fade as they propagate downstream and eventually disappear. Thus this mechanism favors the formation of knots closer to the driving source in agreement with typical observations. Longer variability periods (t >> $\tau_{dy}$) are responsible for the multiple separated bow-shaped shocks observed in some jets (e.g., HH34 and HH 111 jets). Strong support for this conjecture has been given by theoretical studies which have confirmed that traveling shocks created in this way reproduce the essential properties of the observed knots (e.g., Raga et al. 1990, Raga and Kofman 1992, Stone and Norman 1993, de Gouveia Dal Pino 1994, Raga and Canto 1998, de Gouveia Dal Pino 2001).

To simulate a pulsing jet, it is usually applied a sinusoidal velocity variation with time at the jet inlet:

$$v_o(t) = v_j + \Delta v \sin\left(\frac{2\pi t}{P}\right) \quad (6)$$

where $v_j$ is the mean jet speed, $\Delta v$ is the amplitude of the velocity oscillation, and P is its period. Figure depicts fully 3-D hydrodynamical simulations of a radiative cooling pulsed jet showing the development of a chain of internal bow shocks.

Although still a matter of debate, the jet velocity variability that originates the knots could be associated to thermal instabilities in the accretion disk that surrounds the central star, like those believed to be responsible for FU Ori objects (Bell and Lin 1994). Another possibility is that they arise from outbursts produced during violent magnetic reconnection events in the inner edges of the disk (e.g., de Gouveia Dal Pino 2004), or even that both processes occur but at different phases and rates of the accretion process. This subject still deserves further study (see session below).

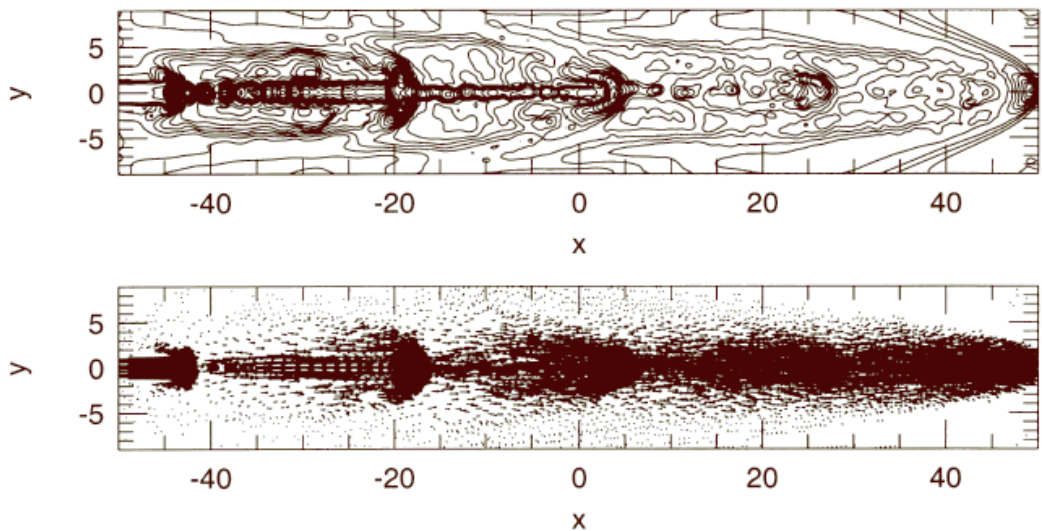

**Figure 4.** Midplane density contour and velocity distribution for a radiative cooling, pulsed jet. The sinusoidal velocity variability at injection has a period ~ 764 yr, a mean jet speed 100 km s$^{-1}$ and amplitude 100 km s$^{-1}$. In the time depicted (~ 4200 yr) the jet has propagated ~0.3 pc. The other initial input parameters are a jet-to-density ratio $\eta = 3$, an ambient density $n_a$= 200 cm$^{-3}$, and a jet radius $R_j = 10^{16}$ cm (from de Gouveia Dal Pino 2001).

*3.3.2. AGN and microquasar jets*

In the case of AGN and microquasar jets, the recent development of multi-dimensional relativistic hydrodynamic codes has allowed, for the first time, the simulation of parsec scale jets and superluminal radio components (e.g., Marti and Muller 2003). The presence of emitting flows at almost the speed of light enhances the importance of relativistic effects in the appearance of these sources. Hence, one should use models which combine relativistic hydrodynamics and synchrotron radiation transfer when comparing with observations.

A variety of processes could produce the string of knots in these classes of jets: nonlinear growth of Kelvin-Helmholtz instabilities, as mentioned before; oblique shock patterns in confined jets; or regular perturbations of the flow velocity or direction. Higher resolution imaging and polarimetry (e.g., with the upgraded VLA) are needed to clarify the nature of this knot phenomenon, which could lead to (model-dependent) estimates of the Mach number in the jets (Bridle 1998).

Linear stability analysis of relativistic flows against Kelvin-Helmholtz perturbations goes back to the seventies. Nowadays, the combination of hydrodynamical simulations and linear stability analysis has provided another step towards the comprehension of relativistic jets both in extragalactic sources and micro-quasars. It is widely accepted that most of the features (even the large amplitude ones) observed in these jets admit an interpretation in terms of the growth of Kelvin-Helmholtz normal modes [e.g., 3C273 (Lobanov and Zensus 2001), and 3C120 jets (Walker et al. 2001)].

On the other hand, where pressure mismatches exist between the jet and the surrounding atmosphere, reconfinement oblique shocks are produced and the energy density enhancement produced downstream from these shocks can give rise to stationary radio knots as observed in many sources.

As in YSO jets, the superluminal components could be produced by triggering small perturbations in the steady jets which propagate at almost the jet flow speed. One example of this is shown in Fig. 5, where a superluminal component is produced from a small variation of the beam flow velocity at the jet inlet.

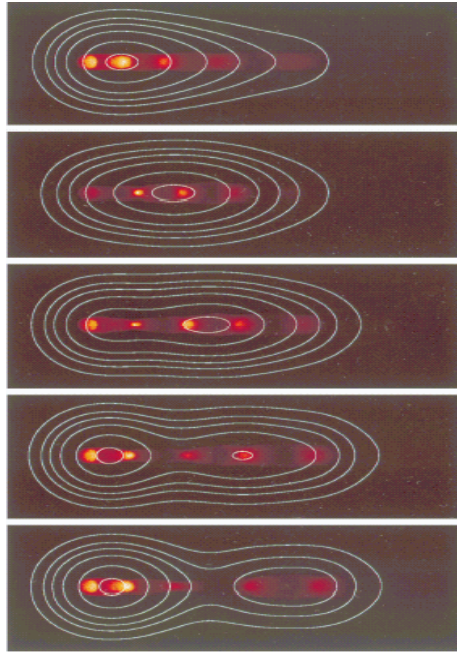

**Figure 5.** Computed radio maps of a compact relativistic jet showing the evolution of a superluminal component with apparent speed 7 times the light speed. Two resolutions are shown: present VLBI resolution (white contours) and resolution provided by the simulation (red images) (from Gómez et al. 1997).

### 3.4. AGN jets composition

In an attempt to solve the basic problem on whether AGN jets are made of proton and relativistic electrons (plasma pair) or instead, of relativistic electron-positron pairs, Scheck et al. (2002) have simulated relativistic $e^-$ p and $e^+$-$e^-$ jets of given kinetic luminosities and jet-ambient density ratios and they found that both *the morphology and the dynamical behavior is almost independent of the assumed jet composition*. Therefore, this remains an open question in the AGN and microquasar jets investigation.

## 4. JET PROPAGATION EFFECTS

### 4.1. Momentum Transfer

It is well known that many young stellar objects are associated not only to the highly collimated, fast YSO optical jets (100-500 km/s) investigated above, but also to less-collimated, slower molecular outflows (< 20 km/s) detected at radio wavelengths. The correlation between both kinds of outflows suggested since the beginning a "unified model" in which the optical jet entrains the molecular gas of the surrounding environment driving the molecular outflow. Determinations of the momentum carried by YSO jets have indicated that they carry, in fact, enough momentum to drive the associated molecular outflows. For example, taking a number density of ions and atoms $n \approx 1.1 \; 10^4$ cm$^{-3}$ in HH 34 jet, and $n \approx 9 \; 10^3$ cm$^{-3}$ in HH 111 jet, Bacciotti et al. (1995) estimated momentum rates $dP/dt = \pi \; R_j^2 \; \rho_j \; v_j^2 \geq 6.7 \; 10^{-5}$ M$_\odot$ yr$^{-1}$ km s$^{-1}$ for HH 34 jet, and $dP/dt \geq$ 1.3 $10^{-4}$ $M_\odot$ $yr^{-1}$ km $s^{-1}$ for HH 111 jet, which are larger than the estimated momentum rates for the associated molecular and neutral outflows: ≈ 5 $10^{-5}$ $M_\odot$ $yr^{-1}$ km $s^{-1}$ and ≈ 1.7 $10^{-5}$ $M_\odot$ $yr^{-1}$ km $s^{-1}$ for the HH 111 and HH 34 outflows, respectively. These determinations have motivated several authors to explore the mechanisms by which the YSO jets can transfer momentum to the molecular outflows.

There is now a general consensus that the molecular emission in YSO jets is mainly due to molecular ambient gas entrained into the atomic flow through the bow shocks (in a *prompt* entrainment, instead of the turbulent shear entrainment that seems to occur in AGN jets of the FRI class, for example) (e.g., Raga et al. 2003, Masson et al. 1993, Raga and Cabrit 1993, Chernin et al. 1994, Downes and Ray 1999). The IR $H_2$ emission of some objects (e.g., HH46, HH47) seems to favor this interpretation, at least qualitatively.

The momentum transfer efficiency through a bow shock, $\varepsilon_{bs}$, which is given by the ratio between the rate of transfer of momentum to the ambient medium through the bow shock and the total input rate of momentum at the jet inlet, can be evaluated analytically and is given by (Chernin et al. 1994):

$$\varepsilon_{bs} \approx 1/\alpha\ [\ 1 + (\eta\alpha)]^2 \tag{7}$$

where $\eta$ and $\alpha$ have been defined in eq. (1).

Investigating the momentum transfer through 3-D numerical simulations of steady-state jets moving in an ambient with Mach numbers in the range $3 < M_a = v_j/c_a < 110$ and density ratios between the jet and the ambient medium $0.3 < \eta < 100$, Chernin et al. (1994) find that the turbulent entrainment along the jet beam is important only for low Mach number ($M_a \leq 6$), low density ratio ($\eta \leq 3$) jets. Although this mechanism can be relevant for AGN jets which seem to fit those values of $M_j$ and $\eta$, it is not the case for the typical YSO jets for which $10 < M_j < 40$, and $\eta \geq 1$. The momentum transfer efficiency found in these numerical models is in agreement with eq. 7 for high Mach number jets indicating that in these cases the (prompt) entrainment at the bow shock is dominant.

### 4.2. Jet precession

It has been found that 5% of the planetary nebulae are point symmetric, namely, their morphology exhibits point reflection symmetry (or an S-shape) about their center (Livio 1998). Similar examples have also been found among extragalactic jets (e.g., NGC 6543 and NGC 5307), and also among the YSO jets (e.g., the giant HH34 jet) (Reipurth and Bally 2001). It has been suggested that the point-symmetry morphology is produced by the episodic ejection of material into a two-sided precessing (or wobbling) jet (e.g., Raga, Canto, and Biro 1993, Livio and Pringle 1996).

In the context of the YSO jets, the existence of precession is generally ascribed to tidal forces produced by a companion in a binary or multiple system. In the case of HH 34, for example, the binary source has not yet been resolved, but there is some evidence that this source could be a binary, such as the discovery of a second outflow (HH 534) emanating from the source, as well as, the abrupt change of direction of the jet axis near knot B (Reipurth et al. 2002). In a recent study, Masciadri et al. (2002; see also de Gouveia Dal Pino 2001) have carried out 3-D hydrodynamical simulations of the full extent of the giant HH34 flow, including not only jet velocity variability to produce the observed chain of bright bow shocks, but also a long period of precession (of the order of $10^4$ yr), which has been able to remarkably reproduce the observed S-shaped morphology of the jet (see Fig. 6). These results are able to reproduce the HH 34 giant flow structure with single precession and variability periods (without the need to invoke different ejection properties at different times of the star lifetime).

In the context of AGN and the other classes of jets, it was not clear until recently, what could cause the jet to precess. An ingredient for the production of a precessing jet in a compact source surrounded by an accretion disk (see Session 6), has been provided by the work of Pringle (1996), who showed that an initially tilted accretion disk which is irradiated by a central source can become unstable to warping (see also Livio 1998). Once the disk is warped, it starts to wobble or precess. It has been shown that radiation induced warping occurs at all radii that satisfy:

$$R/R_S \geq 20\ \delta^2/\ \varepsilon^2 \tag{8}$$

Where $R_S = 2GM/c^2$ is the gravitational radius of the central object, $\delta = \nu_2/\nu_1$ is the ratio of the shear viscosities in the (R,z) and (R,$\phi$) planes of the disk, respectively ($\nu_2$ is the viscosity that acts to reduce the disk tilt), and $\varepsilon = L_*/m_{acc}\ c^2$, where $L_*$ is the luminosity of the central source and is the mass accretion rate $m_{acc}$ into the disk.

Applications of this mechanism to explain point symmetric morphologies in the jets of planetary nebulae, supersoft X-rays stars, and AGNs indicate that warping induced by radiation could be an important mechanism in many classes of objects. Recently, Lai (2004) has investigated the alternative possibility that the warping instability could be driven by magnetic torques associated to the outflow rather than by the radiation, and Romero et al. (2000) have invoked the tidal forces scenario from a black hole companion to explain the precession in some AGN jets, like 3C273, for example (see also Caproni and Abraham 2004).

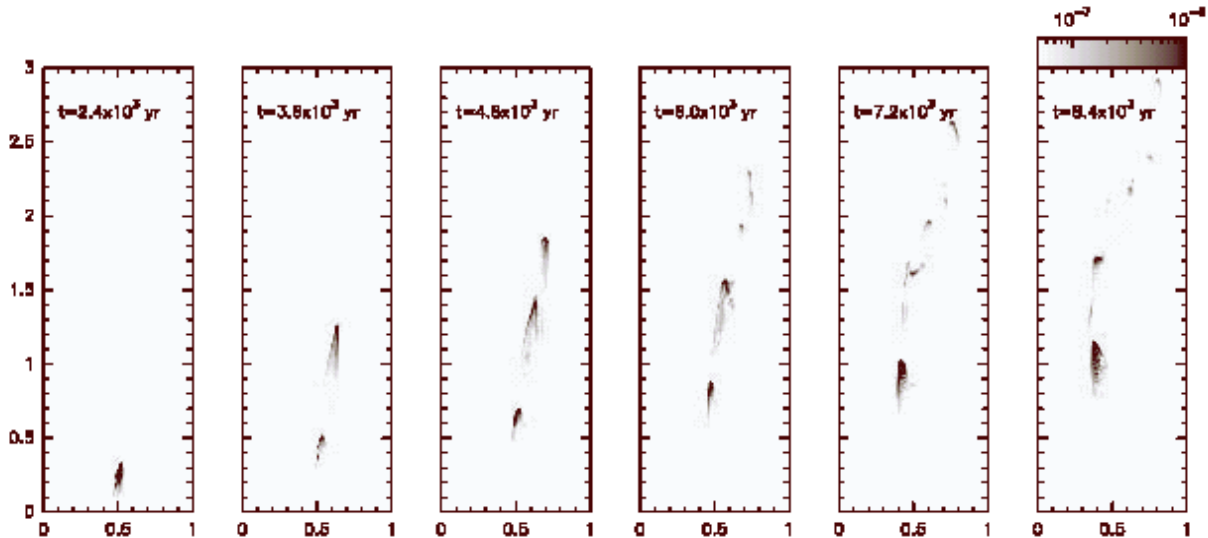

**Figure 6.** Temporal evolution of the Hα maps obtained from a 3-D gas dynamics simulation of the giant HH 34 flow with sinusoidal ejection velocity variability (with a mean velocity of 300 km s$^{-1}$, a half-amplitude of 100 km s$^{-1}$ and a period of 1010yr) and precession of the outflow axis (with a half-angle of 6$^o$ and a 12000 yr period). The abscissas and ordinates are labeled in units of $10^{18}$ cm. The map is shown with the logarithmic gray-scale given (in erg cm$^{-2}$ s$^{-1}$ sr$^{-1}$) by the bar on the last top panel. The similarity with the observed map is remarkable. The maximum Hα intensity is of 2.5 x $10^{-5}$ erg cm$^{-2}$ s$^{-1}$ sr$^{-1}$. The simulation was performed with the adaptive grid code Yguazu (Raga et al. 2000). We note that, from a numerical point of view, it was not a simple task to reproduce the evolution of the outflow over such a large spatial and temporal extent, particularly given the small initial radius of the beam. Indeed, in order to cover the whole domain of 1.5 pc, the computation was carried out on a seven-level, binary adaptive grid with a maximum resolution along the three axes of 1:95 x$10^{15}$ cm (from Masciadri, de Gouveia Dal Pino, Raga, Noriega-Crespo 2002).

### 4.3. Jet deceleraion

An important characteristic of the giant YSO jets is that they appear to slow down for increasing distances from the outflow source. This effect is seen in the HH 34 (Devine et al. 1997) and in the HH 111 giant jets (Reipurth, Bally, and Devine 1997). In a previous study, de Gouveia Dal Pino (2001) modeled giant jets by performing three-dimensional simulations of overdense, radiatively cooling jets modulated with long-period (P of several hundred years) and large amplitude sinusoidal velocity variability, and she found that the multiple traveling knots that develop have their velocity falling off smoothly and systematically with distance. The deceleration was found to be mainly due to progressive momentum transfer sideways into the surrounding medium by the expelled gas from the knots. The more recent study of Masciadri et al. (2002) (see Fig. 6) has found that the impact of the precession on the deceleration mechanism is also considerable. They conclude that the combined effects of jet velocity variability and precession (and the resulting enhanced interaction with the surrounding environment) produce the observed deceleration.

In the case of AGN jets, the dichotomy between FRI and FRII jets (Session 2.2) seems to be due to environmental effects. An indication for that is the fact that more radio power is required to form a FR II radio jet (Owen and Ledlow 1994). Though intrinsic explanations have been hypothesized (see, e.g. Massaglia 2003), it seems that both jet classes are basically similar near the nucleus and that differences in the environment are able to destabilize (possibly via onset of turbulence at the boundary layer between the jet and the ambient medium) and decelerate FR I jets more effectively than FR II jets that succeed to propagate, nearly unchanged, up to the working surface to produce the bright lobe with *hot spots* without much deceleration (Bicknell 1995, Bowman, Leahy, and Komissarov 1996). Another clue that favors this interpretation of the effects of the environment was the observation of six Hybrid MOrphology Radio Sources (HYMORS) that show FR I morphology on one side of the core and FR II morphology on the other side: this is a clear indication that the environment plays a basic role in determining the radio jet appearance at kilo-parsec scales (Gopal-Krishna and Wiita 2000).

The key difference between the two classes may thus be that the fast central "spine" decelerates to sub-relativistic speeds within the galaxy in FRI sources, but persists as far as the distant hot spots in FRII sources (Bridle 1998, Laing and Bridle 2002).

### 4.4. Jet/ambient clouds interactions

Jets emerging from YSOs and AGNs propagate into complex ambient media. YSO jets, in particular, are immersed in molecular clouds which are generally non-homogeneous and formed of many dense clouds. Basically all the models discussed so far have considered the propagation of the jets in initially homogeneous ambient medium. 3-D simulations of cooling jets propagating into more realistic environments with density and pressure stratification have been reported in a number of works (see, e.g., de Gouveia Dal Pino 1999, Raga et al. 2003, for YSO environments). Although these outflows are usually observed to propagate away from their sources in approximately straight lines there are some cases in which the outflow is observed to be deflected (e.g., HH 110 and HH 30 YSO jets, and the molecular outflow in L1221).

Particularly, in the case of the YSO jet HH 110, there is evidence of molecular emission along the jet but with an offset to one side of the atomic jet emission. This has been interpreted as a result of the collision of an YSO atomic jet with a dense molecular cloud (Reipurth, Raga, and Heathcote 1996). As an example, Fig. 7 shows a 3-D hydrodynamical simulation of the impact of a radiative cooling jet with a dense ambient cloud (de Gouveia Dal Pino 1999). The morphology and kinematics of the deflected beam is very similar to that observed in the HH 110 jet. More recent 3-D gas dynamical simulations by Raga et al. (2002) have shown evidence that the molecular emission could in fact come from the dense cloud molecular material that has been ablated by the ionic jet. All these similarities strongly support the proposed jet/cloud interaction interpretation for this system. The morphology of AGNs jets that are observed to bend due to ambient effects has been also investigated with 3-D simulations (see, e.g., Massaglia 2003, and references therein).

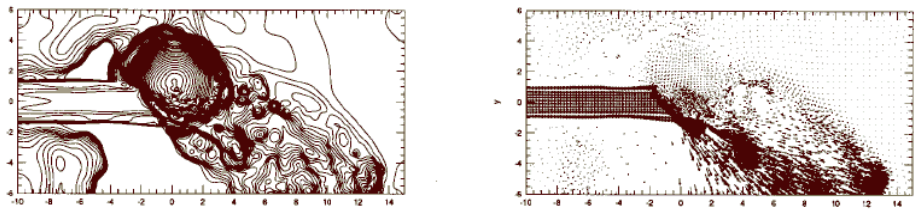

**Figure 7.** Mid-plane density contour (left) and velocity field distribution (right) of a radiatively cooling jet interacting with a cloud with radius twice as large as the jet radius and a density 300 times larger. The initial input parameters are a jet to ambient density ratio $\eta = 3$, $n_a=1000$ cm$^{-3}$, jet radius $R_j = 2 \times 10^{15}$ cm, and $v_j = 200$ km s$^{-1}$ (from de Gouveia Dal Pino 1999).

### 4.5. The End of a Jet

Galaxy clusters are large agglomerations of up to a few thousand galaxies. They also contain large quantities of hot, X-ray emitting gas at T ~ $10^{6-8}$ K (e.g., Bruggen and Kaiser 2002). It has been suggested that the intercluster medium could be heated by AGN jets (e.g., Kaiser and Binney 2003). The energy necessary to heat a galaxy cluster is equivalent to the amount produced by $10^9$ SN explosions and AGN jets can transport from their sources to the ambient medium about 10 times more energy than this ($10^{54}$ J). Jets which are not very powerful (FRI) deposit their energy directly into the surrounding ambient medium via turbulent mixing, while FRII jets deposit most of it into the lobes at the edge far outside the galaxy into the intergalactic or intra-cluster medium. The lifetime of these jets is about $10^{7-8}$ yrs. When they switch-off, the cocoon is filled with hot, diffuse gas, while the surrounding ambient medium is much denser and colder. As a consequence, the gas becomes convectively unstable and buoyance occurs, with the tenous, hot gas interpenetrating the cold component, like air bubbles in water. In fact, a very dim mushroom structure is observed in the cocoon arround the M87 jet that could be due to buoyant effects. Once the mushrooms reach an environment with equal density they stop rising and start spreading over the intergalactic medium.

# 5. EFFECTS OF MAGNETIC FIELDS ON THE JET STRUCTUTRE

An important issue in the investigation of the astrophysical jets, is the requirement of the presence of magnetic fields to explain their origin and collimation through magneto-centrifugal forces associated either with the accretion disk that surrounds the central source, or with the disk-source boundary (see Session 6), but what about the effects of the magnetic fields in the structure and propagation of these jets through the environment?

## 5.1. YSO jets

In the case of the YSO jets, polarization measurements (Ray et al. 1997) have evidenced magnetic field strengths $B \sim 1$ G in the outflow of T Tau S at a distance of few tens of AU[6] from the source, which could imply a $\beta$ plasma ratio, $\beta = p_j/(B^2/8\pi) \sim 10^{-3}$ for a toroidal field configuration (for which the magnetic field decays with the distance r from the jet source as $B \propto r^{-1}$), and $\beta \sim 10^3$, for a longitudinal field (for which the magnetic field decays with the distance from the jet source as $B \propto r^{-2}$), at distances $\sim 0.1$ pc. Since ambipolar diffusion does not seem to be able to significantly dissipate them (e.g., Frank et al. 1998), the figures above suggest that magnetic fields may also play a relevant role on the outer scales of the flow provided that the toroidal component is significant.

In a search for possible signatures of magnetic fields on the large scales of the YSO outflows, several MHD investigations of overdense, radiative cooling jets, have been carried out with the help of multidimensional numerical simulations both in two-dimensions (2-D) (Frank et al.1998, 2000, Gardiner and Frank 2000, Gardiner et al. 2000, Lery and Frank 2000, O'Sullivan and Ray 2000, Stone and Hardee 2000, Volker 1999, Todo et al. 1993) and in three-dimensions (3-D) (de Gouveia Dal Pino and Cerqueira 1996, Cerqueira, de Gouveia Dal Pino and Herant 1997, Cerqueira and de Gouveia Dal Pino, 1999, 2001a, 2001b, 2004; see also de Gouveia Dal Pino and Cerqueira 2002, de Gouveia Dal Pino et al. 2003, Cerqueira and de Gouveia Dal Pino 2003 for reviews).

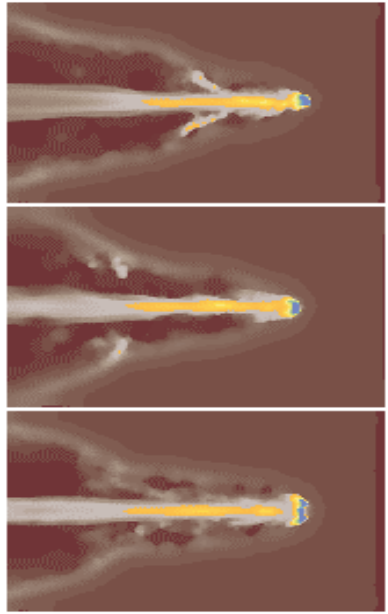

**Figure 8.** Color scale representation of the midplane density of the head of a radiative cooling hydrodynamical jet (top); an MHD jet with initial longitudinal magnetic field configuration (middle); and an MHD jet with initial helical magnetic field configuration (bottom). The initial conditions are: $\eta = n_j/n_a = 3$, $n_a = 200$ cm$^{-3}$, average ambient Mach number $M_a = v_j/c_a = 24$, and $v_j = 398$ The initial $\beta$ for the MHD cases is $\sim 1$. The color scale (from maximum to minimum) is given by black, gray, white, orange, yellow, and blue (from Cerqueira, de Gouveia Dal Pino, and Herant 1997).

The main results of these studies can be summarized as follows. The effects of magnetic fields are dependent on both the field-geometry and intensity (which, unfortunately, are still poorly determined from observations). The presence of a helical or a toroidal field tends to affect more the characteristics of the fluid, compared to the purely HD calculation, than a longitudinal field (see Fig. 8). However, the relative differences that are detected

---

[6] 1AU = one Astronomical Unit = $1.5 \times 10^{13}$ cm (it corresponds to the Earth-Sun distance)

in 2-D simulations involving distinct magnetic field seem to decrease in three-dimensions. In particular, Cerqueira and de Gouveia Dal Pino (2001a, 2001b) have found that features like the nose cones, i.e., elongated structures that often develop at the jet head in 2-D calculations as a consequence of the pinching of the double shock front by toroidal magnetic fields, are absent in the 3-D models, a result that is consistent with observations which show no direct evidence for nose cones at the head of protostellar jets. 3-D calculations have revealed that magnetic fields that are initially nearly in equipartition with the gas (i.e., $\beta \sim 1$) tend to affect only the detailed structure behind the shocks at the head and internal knots, mainly for helical and toroidal topologies. In such cases, the H$\alpha$ emission behind the internal knots can increase by a factor of up to four relative to that in the purely hydrodynamical jet.

The importance of magnetic effects in the HH bow shocks has been also recently reinforced by high resolution studies of individual HH objects. For instance, detailed shock diagnosis of the HH 7 object (Smith, Khanzadyan, and Davis 2003) has revealed that the observed H$_2$ emission is due to a C-shock (where heating is provided by ambipolar diffusion) with a magnetic field strength B $\sim 10^{-4}$ G.

Further 3-D MHD studies are still required as the detailed structure and emission properties of the jets seem to be sensitive to multidimensional effects when magnetic forces are present. Also obvious is the need for further observations and polarization mapping of star formation regions, for a real understanding of their magnetic field structure.

### 5.2. Relativistic Jets

Magnetohydrodynamic simulations of relativistic jets have been used for over a decade to shed light on the physical processes taking place within the sources and to study the implications of ambient magnetic fields in the morphology, propagation and bending properties of AGN and microquasar jets.

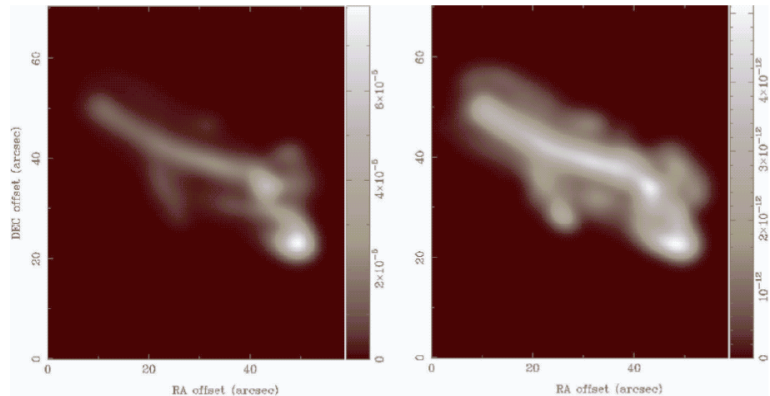

**Figure 9**. Intensity maps constructed from the simulations of MHD relativistic jets as described in the text. The left panel gives the synchrotron emission at 1.4 GHz; the right panel gives the inverse-Compton emission at 10 keV (from Trevillis, Jones, Ryu, Park 2003).

In particular, recent 3-D simulations of AGN jets have been performed that explicitly model the transport of relativistic electrons, including diffuse (Fermi) acceleration at the shocks, as well as, radiative and adiabatic cooling in *light* (underdense) jets, designed to explore the effects of shock acceleration on the nonthermal particles that give rise to synchrotron and inverse-Compton radiations (Tregillis et al. 2003). Because the goal was to explore the connection between the large-scale flow dynamics and the small-scale physics underlying the observed emissions from real radio jets, these studies have combined the magnetic field and relativistic electron momentum distribution information to construct maps of the radio synchrotron and also of the Inverse-Compton emission in X-rays. Fig. 9 illustrates *synthetic* radio synchrotron and X-ray IC images constructed in this way from a 3-D simulation described by Jones, Tregillis and Ryu (2002). Briefly, it represents emission from a light, supersonic jet, with density $10^{-2}$ of the ambient medium, internal Mach number of 8, and a jet velocity of $0.1c$. The jet inflow slowly precesses around a cone of opening angle $5^{o}$, to break cylindrical symmetry. The jet core radius was $r \sim 1$ kpc, and the magnetic field on the jet axis was $B \sim 1$ $\mu$G, with a gas pressure there 100 times greater than the magnetic pressure. This leads to a jet kinetic power of $10^{37}$ W. For this simulation a relativistic electron population with a power law spectrum was brought onto the grid with the jet flow. That population was subjected to acceleration at shocks and adiabatic losses. The 1.4 GHz synchrotron luminosity is 5.3 $10^{32}$ W and the 10 keV inverse Compton luminosity is 6.5 $10^{32}$ W. In the images one can clearly separate the jet, a terminal hot spot, a secondary hot spot and the diffuse lobe.

One of the greatest potential applications of the computation of synthetic maps is a comparison between the real physical properties of a simulated object and observationally inferred source properties, which are generally based on simplified assumptions. For example, from comparison of the synthetic X-ray and synchrotron fluxes of the maps of Fig. 9, Trevillis et al. (2003) infer a magnetic field within the source B ~ 1 µG which is quite consistent with the actual RMS field averaged over the lobe. The simplest minimum energy calculation from the synthetic spectrum also seems to agree with the actual simulated source properties.

## 6. JET PRODUCTION AND COLLIMATION

Although considerable progress has been made toward understanding the jet structure and propagation, no consensus has been reached concerning the basic mechanism for its origin. Can a universal mechanism of acceleration and collimation that operates in all classes be found? The signs are positive at least according to Livio (1998, 2002).

First, there is observational evidence that almost all systems producing jets contain an accretion disk around the central source. This disk is both a source of energy and provides the required axial symmetry. Second, an examination of all classes of objects which produce jets reveals that the ratio of the observed jet velocity to the escape velocity from the central object, $v_j/v_{escape}$, is of order unity, indicating that jets originate from the center of the accretion disk. In the case of YSOs, there is direct observational evidence (in the form of high-resolution images from HST) linking jets to the centers of accretion disks (e.g., the jet HH30; Burrows et al. 1996). In the case of galactic black hole X-ray transients, the source GRS 1915-105 (a microquasar) provides even more evidence for the connection between the jet and the central part of the disk. Multi-wavelength observations suggest a picture in which the inner disk is episodically accreted while ejecting relativistic plasma which subsequently produces infrared and radio flares by synchrotron emission. Therefore, any universal model of jet acceleration and collimation must have these properties: (i) the jet has to originate from the disk (or the source) center; (ii) the jet velocity has to be of the order of the escape (or Keplerian) velocity from the central disk.

### 6.1. Jet production

Several mechanisms have been proposed in the literature (see e.g., de Gouveia Dal Pino 1995; Shibata 2003; Livio 1998, for reviews on these mechanisms). The most promising universal mechanism for jet acceleration and collimation relies on an accretion disk threaded by a perpendicular large-scale magnetic field (Blandford and Payne 1982). The basic idea is that some magnetic flux is in open field lines, which form a certain angle with the disk's surface. Ionized material is forced to follow the field lines. Since these lines are anchored in the disk and rotate with it, material is centrifugally accelerated along the field lines like a bead on a wire, and the lines are wound up by the rotation of the disk [7] (Spruit 1996) (see Fig. 10).

In the context of YSOs, a variant of this magnetocentrifugal disk-driven model has been proposed in which the outflow arises from an interacting star-disk system (Shu et al. 1994). It assumes a viscous and imperfectly conducting disk (with large magnetic diffusivity) steadily accreting (with small accretion rate $dM_D/dt$) onto a young star with a strong magnetic field. Shielding currents in the surface layers of the disk will prevent stellar field lines from penetrating the disk everywhere except for a range of radii around $r = R_X$, where the Keplerian angular speed of rotation $\Omega_X$ equals the angular speed of the star $\Omega_S$. For the low disk accretion rate and high magnetic fields associated with typical young stars, $R_X$ exceeds the radius of the star $R_S$ by a factor of a few, and the inner disk region is effectively truncated at a radius $R_t < R_X$. Between $R_t$ and $R_X$, the closed field lines bow sufficiently inward and the accreting gas attaches to the field and is funneled dynamically down the effective potential (gravitational plus centrifugal) onto the star. The associated magnetic torques to this accreting gas may transfer angular momentum mostly to the disk. Thus, the star can spin slowly as long as $R_X$ remains $> R_S$. For r

---

[7] When the centrifugal force component along the line exceeds that of gravity, the gas tied to the field line is accelerated outward. This outward centrifugal acceleration continues up to the Alfvén point [the location where the outflow poloidal speed reaches the Alfvén speed $v_A = B_p/(4\pi\rho)^{1/2}$]. Beyond this point ($v \rho^2 > B^2/8\pi$), the inertia of the gas causes it to lag behind the rotation of the field line and the field winds up thus developing a strong toroidal component ($B_\phi$).

> $R_X$ field lines threading the disk bow outward making the gas off the midplane to rotate at super-Keplerian velocities. This combination drives a magnetocentrifugal wind mostly arising from the region around the co-rotation radius $R_X$ with a mass-loss rate $dM_W/dt$ equal to a definite fraction of the disk accretion rate $dM_D/dt$. A schematic diagram of this model is presented in Fig. 11 and a quantitative description is presented in Shu et al. (1994, 1999). They find that stellar magnetic fields of few kilo-gauss can drive outflows with mass-loss rates of $10^{-6}$ $M_\odot$ yr$^{-1}$ from rapidly accreting YSOs (rotating near breakup) and $10^{-8}$ $M_\odot$ yr$^{-1}$ from slowly accreting (slowly rotating) young stars. The mechanism described can accelerate outflows from these systems to the observed velocities $\approx$ few 100 km yr$^{-1}$ within few stellar radii $\leq 10$ $R_S$.

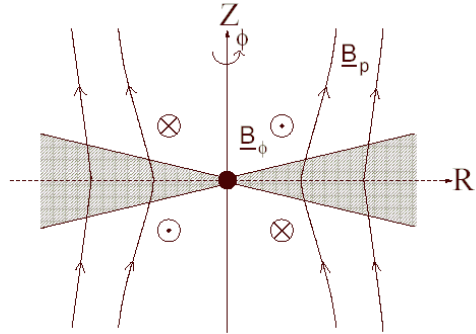

**Figure 10.** Schematic view of the magnetocentrifugal disk-driven model. The accretion disk is shown in cross-section (shaded) in orbit around the central object. The magnetic field has become advected towards the center along with the accretion flow (from Archibald 2004).

More generally, these mechanisms are believed to be universal in the production of bipolar outflows from a wide range of astrophysical objects. In the case of the AGN jets, also there is evidence for the presence of dusty molecular disks near the central nucleus (e.g., Phinney 1989) which gives support to disk-driven outflow models or the combined source-disk-driven outflow models.

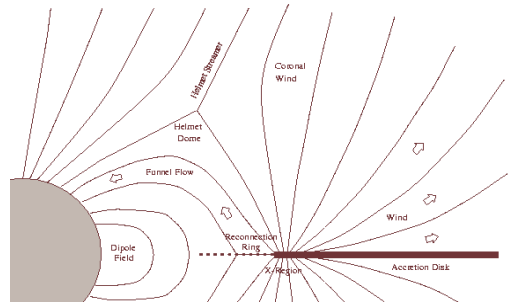

**Figure 11.** Schematic representation of the main components of a magnetocentrifugally driven outflow from a disk-star system. Gas interior to the X-region ($r < R_X$) diffuses onto field lines that bow inward and is funneled onto the star. Gas at $r > R_X$ diffuses onto field lines that bow outward and launch a magnetocentrifugally driven wind (from Shu et al. 1999, see also de Gouveia Dal Pino and Lazarian 2000, 2001).

Since the outflows are in principle capable of transporting the excess of angular momentum of the accreting matter as well as most of the gravitational energy that is liberated, one may attribute the ubiquity of bipolar outflows to the fact that the winds are a necessary ingredient in the accretion process in that they carry away the angular momentum that needs to be removed in order for the accretion to proceed.

### 6.2. Jet Collimation

The collimation process in the magnetocentrifugal models discussed above may occur either by the curvature force ($B_\phi^2/4\pi r$, where r is the radial distance from the disk) exerted by the toroidal magnetic field on the outflowing material (collimation by *hoop* stresses), or by the poloidal magnetic flux which increases with radial distance in the disk (e.g., Livio 1998) (poloidal collimation) if the disk's radius is large relative to that of its central object. The latter process avoids the magnetic kink instability (e.g., Parker 1979) to which hoop stresses collimation is vulnerable.

Both analytic (e.g., Parker 1979) and numerical MHD models in two and three dimensions [e.g., Kudoh, Matsumoto, and Shibata 2002, Ouyed, Clark and Pudritz 2003, Shibata 2003] of magnetized jets are now capable of producing jet-like outflows from the disk center (see Figure 12).

The terminal jet velocities that are obtained from these models satisfy the following relations [e.g., Kudoh and Shibata 1997, Shibata 2003]:

$$v_j/v_K \propto E_{mag}^{1/3} \quad \text{for large } E_{mag}, \quad B_\phi/B_p \ll 1$$

or

$$v_j/v_K \propto E_{mag}^{1/6} \quad \text{for small } E_{mag}, \quad B_\phi/B_p \gg 1$$

Here, $E_{mag} = (v_A/v_K)^2$ = (magnetic energy)/ (gravitational energy), $B_\phi$ and $B_p$ are the initial toroidal and poloidal magnetic field components, respectively, $v_A$ is the poloidal Alfvén velocity [$v_A = (B_p^2/4\pi\rho)^{1/2}$, where $\rho$ is the gas density], and $v_K$ is the Keplerian velocity at the footpoint [$v_K = (MG/d)^{1/2}$, where M is the mass of the central source and d the distance from it]. The important consequence of the equations above is the fact that for a wide range of poloidal magnetic field strengths, the jet velocity is of the order of the Keplerian velocity at the footpoint of the jet, as observed. It is also found that the acceleration is mainly centrifugal for large $E_{mag}$, but that magnetic pressure plays an important role for small $E_{mag}$.

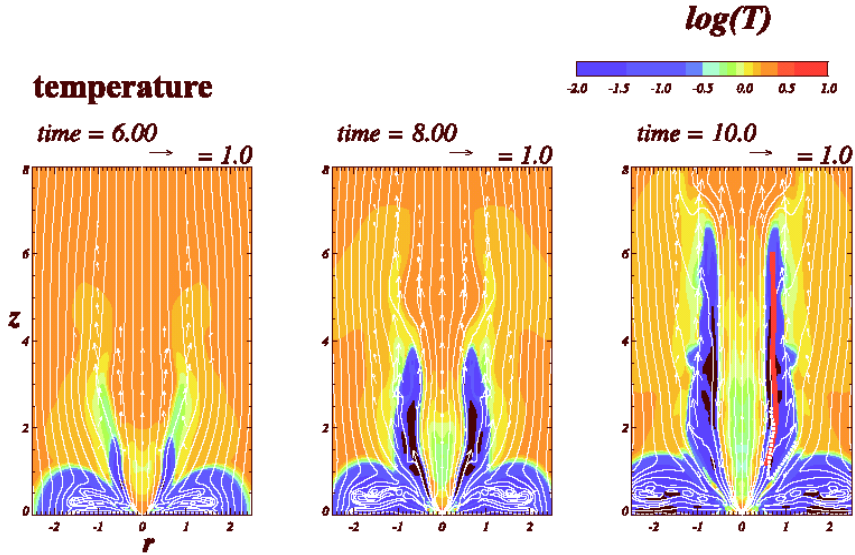

**Figure 12.** Typical example of 2.5D MHD numerical simulations of the formation of MHD jets from a thick disk. The white lines show the poloidal magnetic field and the arrows show the poloidal velocity. The thick red line shows a part of a stream line and the dotted line shows the Alfven surface where the poloidal velocity equals the poloidal Alfven velocity. The disk becomes turbulent because of magneto-rotational instability. In a recent development of accretion disk theory, Balbus and Hawley (1991) showed that perturbations in a differentially rotating magnetized disk grow when the magnetic pressure is smaller than the gas pressure in the disk. This generates a magneto-rotational instability in the disk that can be the origin of the viscosity in it, which in turn can provide an efficient transport of angular momentum in the disk. The simulation above shows that a magnetically driven jet is able to develop even from a disk that is subject to this magneto-driven rotational instability. The instability grows in the disk and causes violent accretion. As it continues, the accretion flow is partially turned outward to the outflow that is accelerated along the poloidal magnetic field lines (from Kudoh, Matsumoto and Shibata 2002).

The poloidal collimation mechanism also predicts the minimum opening angle of the jet (Spruit 1996). For a vertical field with a radial dependence of the form $B_z \sim (r/R_{in})^{-1}$, where $R_{in}$ is the inner radius of the disk, the jet opening angle is given by (e.g. Matsumoto and Shibata 1997)

$$\theta_{min} \approx (R_{in}/R_{out})^{1/2}$$

Where $R_{out}$ is the outer disk radius. An examination of the expected values of $\theta_{min}$ for the different classes of objects which produce jets reveals that, while the values for AGNs, YSOs, the symbiotic star R Aqr, and BHXRTs are of the order 0.01, and thus consistent with the existence of highly collimated jets in these systems, the values for cataclysmic variables and planetary nebulae are of the order 0.1 or larger. Based on this, one could conclude that highly collimated jets should not exist in the latter systems, contrary to recent observational indications. This maybe an indication that additional physical ingredients are necessary for the formation of jets (Livio 1998, 2002).

### 6.3. Open Questions

There are observational facts (other than that discussed in the previous paragraph) that hint the possibility that an accretion disk threaded by a magnetic field may be not sufficient, under some circumstances, for the production of powerful jets alone (Livio 1998, 2002). For example, only two of the BHXRTs (1655-40 and 1915+105) have been observed to produce jets. Similarly, only one short period cataclysmic variable (T Pyx) has been observed to produce a jet. This suggests that powerful jets may require an additional energy/wind source besides the magnetized accretion disk.

Livio (1998) has suggested that in the case of the planetary nebulae and the cataclysmic variable T Pyx, nuclear burning on the surface of the accreting object could provide the extra energy/wind source for these systems.

In the case of the systems with black hole accretors, Livio suggests that the spin of the black hole could provide the extra energy source since rotational energy could be extracted from the spinning black hole (Blandford and Znajek 1997). The mechanism itself can be understood if the black hole is compared to a resistor rotating in a magnetic field and generating a potential difference between the hole's pole and equator.

Zhang, Cui and Chen (1997), in an attempt to determine the spin of the black holes in the microquasars, found that the spin of the sources 1124+68, 2000+25, and LMCX-3 is null, while for 1655-40 and 1915+105, the only BHXRTs with powerful jets, they found a = 0.93 and 0.998, respectively [where $a = J/(GM^2/c)$ is the dimensionless specific angular momentum of the black hole]. This result is consistent with the suggestion that the spin could be a necessary ingredient to produce powerful jets in these systems.

Finally, another intriguing fact is that jets from the nuclei of 'blazars' — active galaxies from which jets point in the direction of the observer — are ultrarelativistic, whereas jets from Seyfert galaxies are not. Yet in both cases, jets are thought to originate from the centers of disks around supermassive black holes. The difference may be due to different environments, different mass-loading of the magnetic field lines, or something else, but it could be also attributed to differences in extraction of rotational energy from the spinning black hole by the magnetic fields. X-ray observations of the Seyfert galaxy MCG-6-30-15 and the galactic black hole candidate XTE J1650-500 have now revealed an extremely broad and redshifted iron $K\alpha$ line, indicating an origin in the very center of the accretion disk. This implied emissivity may mean that energy is indeed extracted from the spinning black hole. If true, this could show that the same exotic process is operating around the two classes of black holes across a range of a factor of a million in black-hole mass!

This interpretation is far from unique, however and other extra energy/wind sources may be operating like the presence of a hot coronae over the disk, and supercritical accretion.